# Molecular polygons probe the role of intramolecular strain in the photophysics of π-conjugated chromophores

P. Wilhelm, J. Vogelsang, G. Poluektov, N. Schönfelder, T. J. Keller, S.-S. Jester, S. Höger, and J. M. Lupton

**Abstract:** ∏-conjugated segments – chromophores – constitute the electronically active units of polymer materials used in organic electronics. To elucidate the effect of bending of these linear moieties on elementary electronic properties such as luminescence colour and radiative rate we introduce a series of molecular polygons. The π-system in these molecules becomes so distorted in bichromophores (digons) that these absorb and emit light of arbitrary polarisation: any part of the chain absorbs and emits radiation with equal probability. Bending leads to a cancellation of transition dipole moment (TDM), increasing excited-state lifetime. Simultaneously, fluorescence shifts to the red as radiative transitions require mixing of the excited state with vibrational modes. However, strain can become so large that excited-state localisation on shorter units of the chain occurs, compensating TDM cancellation. Since these effects counteract, underlying correlations between shape and photophysics can only be resolved in single molecules.

Microscopic molecular geometry can be crucial in determining macroscopic performance of materials in devices such as organic light-emitting diodes (OLEDs). Recently, for example, spontaneous ordering of the molecular emitters in the plane of OLEDs was identified to counteract deleterious optical waveguiding of the emitted light,[1] so that OLED efficiency is now primarily limited by molecular orientation rather than elementary charge carrier recombination kinetics. A powerful but underutilized technique to uncover information on such microscopic structure is single-molecule fluorescence spectroscopy. This approach has revealed information on spontaneous microscopic ordering of π-conjugated macromolecules such as conjugated polymers,[2] and uncovered some elementary interaction pathways between injected charges and excited states.[2g] But a central question has remained hard to address: what is the role of shape – bending and twisting – of the underlying π-conjugated chromophore as it interacts with its environment in space?[2h] A material deposited by thermal evaporation, doctor blading or spin coating inevitably adopts a non-equilibrium conformation,[3] which will impact electronic properties. But as conformational information is extracted from spectroscopy, a direct correlation between microscopic structure and electronic function is often flawed: for example, merely studying polarisation anisotropy[2] does not allow one to ascertain whether a large molecule contains multiple linear chromophores of different orientations, or whether the chromophores themselves are bent. In addition, interchromophoric interactions can impact spectroscopic observables, such as through the formation of H- or J-aggregate excited-state species,[4] making it challenging to differentiate aggregation[5] from effects like twisting or bending arising from strain. Figure 1 illustrates the main effects that bending is expected to have on spectroscopic observables. Bending will increase nuclear relaxation in the excited state, shifting the emission to the red and raising the intensity of vibronic transitions. However, it may also weaken conjugation, shifting absorption to the blue (Figure 1a).[6] Bending will affect the overall transition dipole moment (TDM) by the cancellation of individual monomeric moments, illustrated in panel b). Such a cancellation redistributes oscillator strength from purely electronic to vibronic modes, raising fluorescence lifetime.[4,6] However, localisation of the exciton can occur within the conjugated segment (panel c),[7] for example by twisting of phenyl rings with respect to each other.[8] Such localisation can arise at a different part of the molecule every time a photon is absorbed, which results in unpolarised luminescence[8a]. Localisation will therefore counteract the cancellation effect of TDMs in bent structures, raising oscillator strength and shifting emission to the blue compared to the delocalised excitation in a bent chromophore (panel b), while retaining unpolarised luminescence. Despite solid theoretical backing,[6] it is hard to predict which of these effects will dominate in a given system. Slight conformational variations are anticipated to lead to scattering of observable characteristics on the single-molecule level.

To address this problem, we introduce a series of cyclo-oligomers consisting of $n$ π-conjugated units, based on phenylene-ethynylene-butadiynylenes, connected by apex moieties to yield molecular polygons.[9] The chemical structures are shown in Figure 2 ($n$=4: **4**; $n$=3: **3**; $n$=2: **2** and **2'**). These structures can be considered as cut-outs of the respective conjugated polymers with defined chromophore length and tuneable curvature. The strain on the digon **2** is controlled by the opening angle of the apex units shown in panel b). Details of synthesis and characterisation are given in the Supporting Information. The polygon shapes are clearly resolved in scanning tunnelling microscopy (STM) of self-assembled monolayers of the molecules on graphite, shown in Figure 2c). Although the substrate imposes some distortion, the effect of polygon order on bending of the π-system is apparent: the fewer units are linked, the greater the strain. The carbazole apex has a smaller opening angle of 85° compared to the bithiophene unit (101°), and therefore appears to induce less distortion of the chromophore.

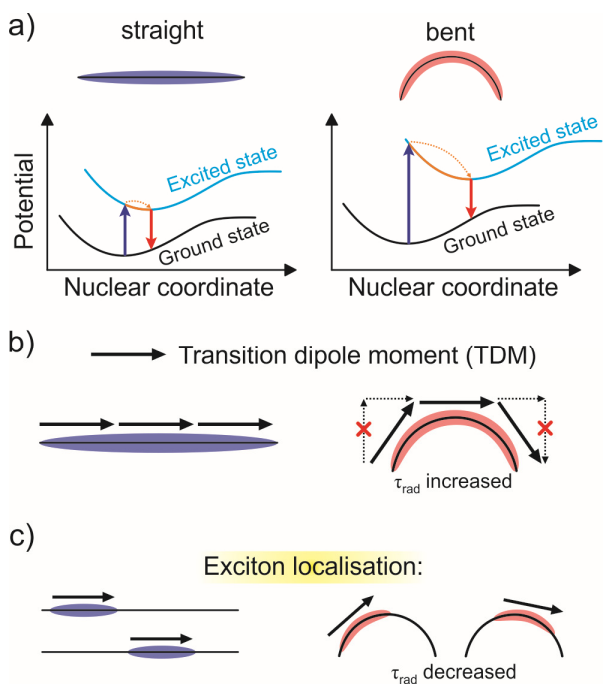

***Figure 1.*** *Influence of bending on optoelectronic properties of π-conjugated chromophores. a) Bending of the π-system promotes structural relaxation in the excited state, increasing Stokes' shift and vibronic coupling in PL. b) In a linear system, TDMs of the individual segments add up. In a bent system, they cancel out, raising the radiative transition lifetime $\tau_{rad}$. c) Exciton localisation to a smaller region of the segment can also occur, which will affect TDM orientation in bent (right) but not in linear (left) chromophores. Localisation in a bent system will therefore counteract TDM cancellation, lowering $\tau_{rad}$.*

We begin by comparing ensemble absorption and photoluminescence (PL) spectra of the compounds in dilute toluene solution in Figure 3. The compounds show fluorescence quantum yields of 55-70 %. Since the chromophores in **4** are least bent, we use these as the reference,[10] with the spectra marked in grey. Increasing bending in **3** and **2** induces a blue shift in absorption (Figure 1a). This hypsochromic shift presumably arises from increased structural disorder of the molecules, which effectively shortens the chromophores.[6] The change in spectral position coincides with a redistribution of oscillator strength to the vibronic transitions since the electronic (0-0) transition becomes forbidden in a perfectly circular system.[8f, 11] Redistribution of oscillator strength from electronic to vibronic transitions is also apparent in the PL spectra. At the same time, the radiative fluorescence lifetime $\tau_{rad}$, the ratio of PL lifetime to quantum yield, almost doubles between square **4** and digon **2**. **3** and **2** show no spectral shift in emission compared to **4**. An increase in nuclear rearrangement of bent structures provides an explanation for this observation (Figure 1a). In contrast, the carbazole-clamped digon **2′** shows a blue shift of PL and absorption because of the inductive effect of the polar carbazole unit, but does not exhibit much change in vibronic to electronic ratio of the PL spectrum compared to **4**. $\tau_{rad}$ is decreased in **2′** compared to **2**. These ensemble-based observations suggest a link between molecular shape, and PL spectral characteristics and fluorescence lifetime.

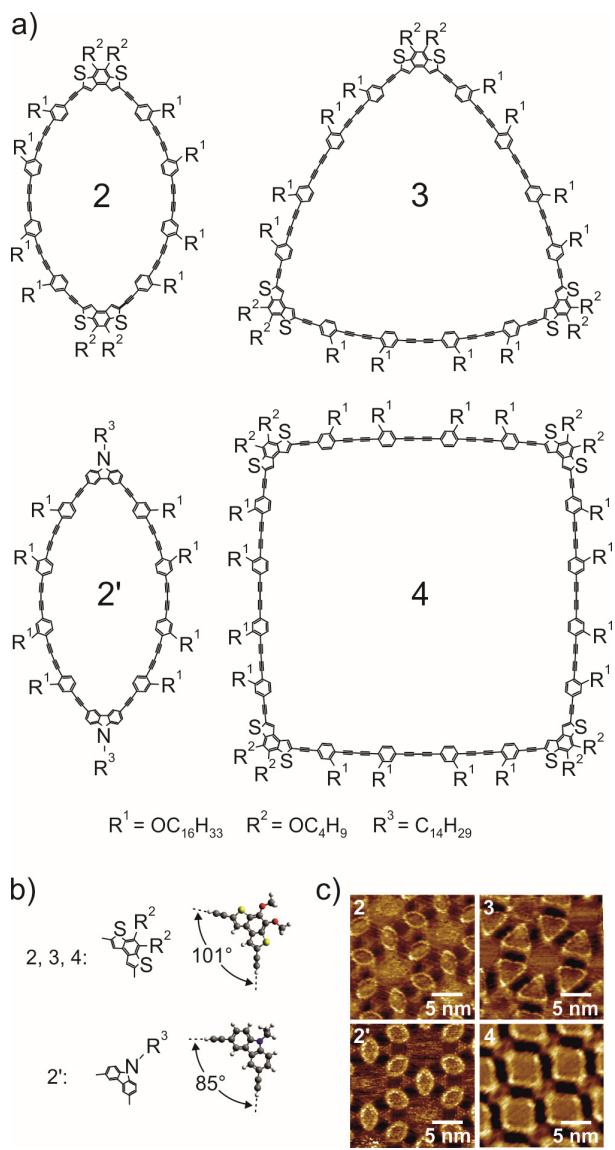

*Figure 2. Molecular polygons probe the influence of chromophore bending on electronic properties. a) Phenylene-ethynylene-butadiynylene-based digons (**2**), triangles (**3**) and squares (**4**), with (b) bithiophene or carbazole clamping units. c) STM images of the samples on a graphite surface.*

To assess the degree of bending of the individual chromophores we consider microscopic TDM orientation. This can be assessed by performing polarisation-resolved absorption (*A*) and emission (*E*) spectroscopy on the single-molecule level, with the molecules embedded in a PMMA matrix film on a glass slide as described previously.[8a] As sketched in Figure 4, the PL intensity *I* is measured for excitation with both horizontally (*H*) and vertically (*V*) polarised light, and the fluorescence is also resolved into *H* and *V* components. The ratio between difference and sum of the two intensities gives the linear dichroism *LD*, either in absorption or emission. With no preferential orientation of the molecules in space, the TDM orientation with respect to light polarisation is not fixed. A random distribution of different molecular orientations must therefore be considered. However, for comparably large macrocycles, we recently reported evidence that the molecules tend to preferentially align in the plane of the film due to the forces induced by the embedding matrix.[8f] For such a quasi-two-dimensional orientation of the TDM in a plane vertical to the microscope axis, a well-defined TDM alignment along the *H* or *V* orientation of the arbitrary laser or detector polarisation will give maxima at ±1 in the distribution histograms,[8a,f] whereas an unpolarised absorber or emitter will only yield LD values of 0. The squares **4** yield, on average, unpolarised absorption since they contain two pairs of chromophores with orthogonal TDMs whose net orientation effectively cancel out in the measurement. Interestingly, the emission is also unpolarised. This fluorescence isotropy can be a consequence of the orthogonal chromophores emitting independently of each other. Alternatively, one may consider a fluctuation in energy transfer pathway between the chromophores so that a different chromophore emits for each excitation event,[12] or singlet-singlet annihilation of excited states between the chromophores. As discussed in the Supporting Information, the latter is the case since all compounds show photon antibunching, the deterministic emission of one single photon from the multichromophoric molecule at a time (see Figure S2). This observation implies that fluorescence arises non-deterministically from one of the four chromophores, giving effectively unpolarised emission, i.e. the source of the fluorescence switches between the different TDM orientations in the molecule. The $LD_A$ histogram of the triangle is identical to that of the square. The $LD_E$ histogram even appears slightly narrower than for **4**. The surprise comes from the digon **2**, with two nominally parallel chromophores. The $LD_A$ histogram is indistinguishable from those of **3** and **4**, and the tails of $LD_E$ appear to drop off even faster. The opposite is true in the digon with the narrower opening **2´**. Here, both LD distributions approach those expected for a single TDM of arbitrary orientation,[8a] even though the STM images suggest that **2** and **2´** have similar conformations. In STM, the π-system can appear distorted due to interactions between neighbouring molecules and the underlying substrate lattice. Fluorescence measurements of LD, in contrast, report directly on the range of conformations which can be adopted in a bulk film. We conclude that **2** is more strongly bent than **2´**, giving rise to virtually unpolarised excitation in **2**. The resulting histogram in emission can only arise if the exciton localises within a chromophore on the molecule and if this localisation fluctuates with time, leading to emission which is overall unpolarised.

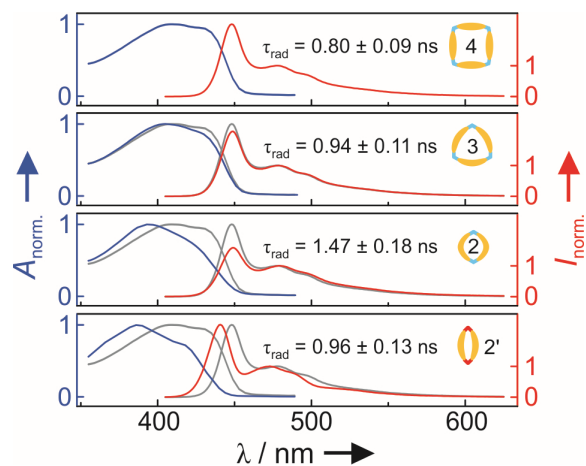

*Figure 3. Ensemble absorption and emission spectra of the compounds, compared to the spectra of the square with the least distorted chromophores (grey). The radiative lifetime is stated for each material.*

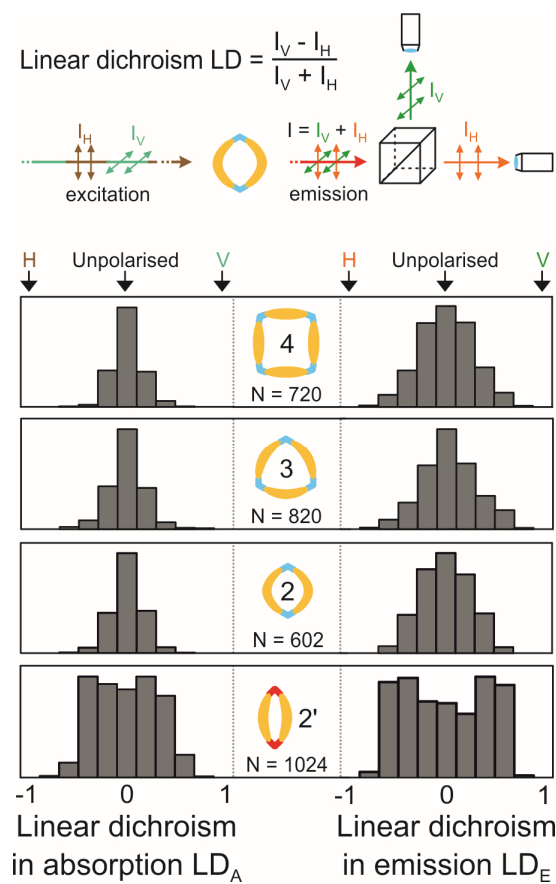

*Figure 4.* Linear dichroism histograms for N molecules in absorption and emission of single molecular polygons in a PMMA matrix. An unpolarised absorber or emitter will lead to a peak in the histogram at 0, whereas an H or V polarised unit will result in peaks at ±1.

As discussed in Figure 1, bending of the emitting chromophore should increase the radiative lifetime $\tau_{rad}$ due to a cancellation of oscillator strength, and should impact the emission spectrum.[4,6] Because the PL quantum yield is similar for all compounds, a change of $\tau_{rad}$ will induce a corresponding modification of PL lifetime $\tau_{PL}$.[13] However, since ensemble spectroscopic data are inhomogeneously broadened over all conformations, they only offer limited insight into microscopic structure. Figure 5a) correlates the PL position of the electronic transition, $\lambda_{Peak}$, and $\tau_{PL}$. For all compounds $\lambda_{Peak}$ scatters over 20 nm because of slight changes in chromophore shape and interactions with the embedding matrix.[15] A scatter is also seen in $\tau_{PL}$, the magnitude of which depends on the molecule. For **4**, there is no discernible correlation between $\lambda_{Peak}$ and $\tau_{PL}$ (see Figure S3). For **3**, a slight correlation emerges in the plot, even though the average of $\lambda_{Peak}$ remains unchanged. The correlation becomes clear in **2**, with $\tau_{PL}$ scattering by 250%, and can be explained by variations in the degree of excited-state nuclear relaxation of the individual molecules. In bent molecules where excited-state reorganization enhances conjugation by reducing twisting, the spectrum will red-shift with a concomitant increase of $\tau_{PL}$. Molecules with short chromophores will have shorter $\tau_{PL}$ and emit at shorter wavelengths (Figure 1b). We therefore group the PL spectra of **2** by PL lifetime as shown in panel b). As $\tau_{PL}$ increases, spectral intensity is redistributed to the vibronic around 470 nm as would be expected for an increase of digon curvature (cf. Figure 1a). In contrast, in the less deformed **2′**, the $\lambda_{Peak}$–$\tau_{PL}$ correlation is barely discernible.[14]

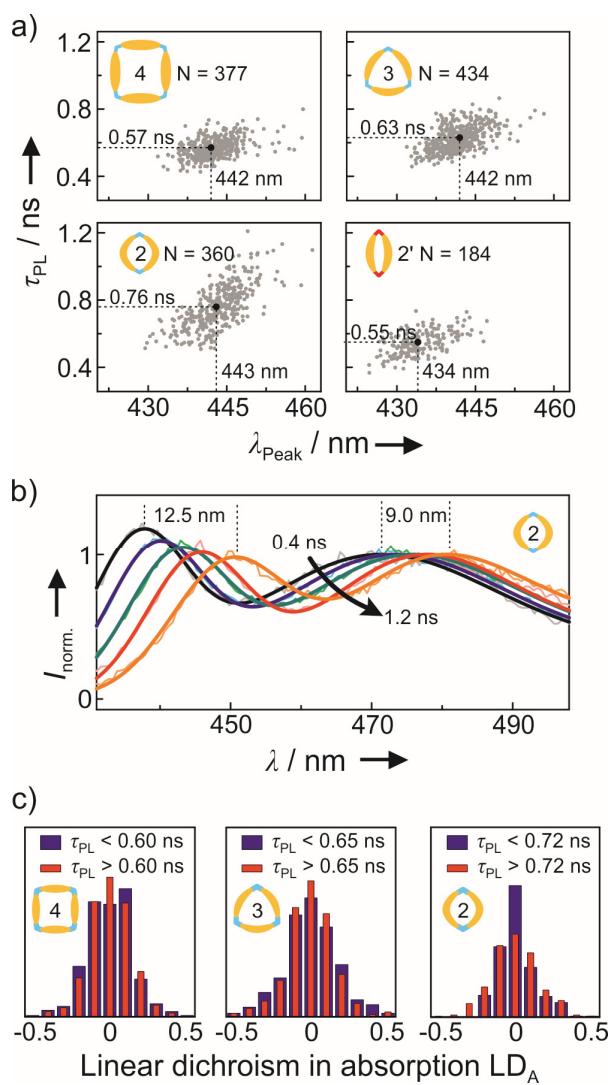

*Figure 5. Effect of chromophore bending on PL lifetime and spectrum. a) Scatter plots of PL lifetime and PL peak wavelength of the electronic transition. The average values are stated along with the sample size N. b) Plot of the averaged emission spectra from panel a) for compound **2**, selected by $\tau_{PL}$ (black: 0.4-0.6 ns; blue: 0.6-0.7 ns; green: 0.7-0.8 ns; red: 0.8-0.9 ns; orange: 0.9-1.2 ns). All spectra were shifted to the average central*

*wavelength for a particular time window to remove the effect of random spectral scatter between molecules. c) $LD_A$ histograms selected by the 50 % shortest and longest-lived molecules. Only **2** shows a difference with the shorter-lived emitters appearing most bent, showing overall lower $LD_A$ values. Random exciton localisation must therefore occur after every photoexcitation event on different parts of the bent chromophore to prevent the TDM cancellation sketched in Fig. 1b).*

Crucially, the average of $\lambda_{Peak}$ is virtually unchanged for **2**, **3** and **4**, suggesting similar effective chromophore sizes: averaging masks the effect of bending on the photophysics. To address this counterintuitive effect, we examine the degree of bending of the single molecule, as given by the $LD_A$ distribution, and relate this to $\tau_{PL}$ (and therefore, in the case of **2**, $\lambda_{Peak}$). Figure 5c) plots $LD_A$ for the 50% highest and lowest values of $\tau_{PL}$. No difference is seen for **4** and **3**, but for **2**, short lifetimes correlate with lower $LD_A$ values, implying more bending of the chromophores exhibiting shorter lifetime. This effect is exactly the opposite of what would be expected from Figure 1b). The observations can only be rationalised if strong bending promotes exciton localisation by reducing the effective conjugation in the excited state (Figure 1c), leading to shorter chromophores with bluer emission (Figure 5b) and lower $\tau_{PL}$ (Figure 5c).[6] Such localisation can reduce the effect of TDM reduction and the corresponding increase in $\tau_{PL}$ arising from bending (Figure 1b), but will still lead to unpolarised fluorescence ($LD_E \rightarrow 0$), provided that localisation occurs randomly on different parts of the bent segment with each single photoexcitation event. We conclude that as bending increases, most likely twisting of the phenyl rings arises, promoting exciton localisation.[8c-d] As a consequence, the pronounced $\lambda_{Peak}$–$\tau_{PL}$ correlation of the digon in Figure 5a) emerges from the heterogeneity of the degree of exciton localisation between single molecules. These polygons therefore allow a quantitative distinction between the effect of bending and torsion-induced excited-state localisation. Such degrees of freedom have yet to be exploited in the design of materials for specific applications, such as OLED emitters with desired polarization directionality.

**Acknowledgements**

We are indebted to the European Research Council for funding through the Starting Grant MolMesON (305020), the DFG (SFB 813 and GRK 1570) and to the Volkswagen Foundation for continued support of the collaboration.

**Keywords:** Single-molecule spectroscopy • macrocycles • Intramolecular strain • Organic electronics